\documentclass[conference]{IEEEtran}

\usepackage{graphicx}
\usepackage{amsmath, amsthm, amssymb,textcomp}

\usepackage{algorithm}
\usepackage{cite}
\usepackage{booktabs}

\usepackage{url}
\usepackage{subcaption}

\usepackage{algorithmicx}
\usepackage{algpseudocode}
\usepackage{paralist}

\usepackage{color}

\usepackage{multirow}
\begin{document}
%
% paper title
% can use linebreaks \\ within to get better formatting as desired
\title{\Huge{Homomorphic Data Isolation for Hardware \\ Trojan Protection}}

%\author{ M. Tarek Ibn Ziad, Amr Alanwar, Yousra Alkabani, M. Watheq El-Kharashi, %Hassan Bedour\\
%Department of Computer and Systems Engineering, Ain Shams University\\
%Cairo 11517, Egypt\\
%Department of Electrical Engineering, University of California Los Angeles, %USA\\
%\{mohamed.tarek, amr.alanwar, yousra.alkabani, watheq.elkharashi, %hassan.bedour\}@eng.asu.edu.eg}
%alanwar@ucla.edu

\author{\IEEEauthorblockN{M. Tarek Ibn Ziad\IEEEauthorrefmark{1},
Amr Alanwar\IEEEauthorrefmark{2},
Yousra Alkabani\IEEEauthorrefmark{1}, 
M. Watheq El-Kharashi\IEEEauthorrefmark{1} and
Hassan Bedour\IEEEauthorrefmark{1}}
\IEEEauthorblockA{\IEEEauthorrefmark{1}Department of Computer and Systems Engineering, Ain Shams University, Cairo, Egypt}
\IEEEauthorblockA{\IEEEauthorrefmark{2}Department of Electrical Engineering, UCLA, Los Angeles, CA, USA}
Email: mohamed.tarek@eng.asu.edu.eg, alanwar@ucla.edu, \\ \{yousra.alkabani, watheq.elkharashi, hassan.bedour\}@eng.asu.edu.eg}

%\author{
%\\
%\\
%\\
%\\
%}
\maketitle

\vspace{10pt}
\begin{abstract}
%\boldmath
The interest in homomorphic encryption/decryption is increasing due to its excellent security properties and operating facilities. It allows operating on data without revealing its content. In this work, we suggest using homomorphism for Hardware Trojan protection. We implement two partial homomorphic designs based on ElGamal encryption/decryption scheme. The first design is a multiplicative homomorphic, whereas the second one is an additive homomorphic. We implement the proposed designs on a low-cost Xilinx Spartan-6 FPGA. Area utilization, delay, and power consumption are reported for both designs. Furthermore, we introduce a dual-circuit design that combines the two earlier designs using resource sharing in order to have minimum area cost. Experimental results show that our dual-circuit design saves 35\% of the logic resources compared to a regular design without resource sharing. The saving in power consumption is 20\%, whereas the number of cycles needed remains almost the same. 
\\ \\ 
\small\textbf{Keywords- ElGamal Encryption, Hardware Trojan, Homomorphism, Security}

\end{abstract}

%\category{ESS3}{Embedded System Validation, Verification, Security, Dependability}{Hardware and software security techniques}

%\terms{Theory}

%\keywords{Hardware Trojan, Trojan Triggering}
% IEEEtran.cls defaults to using nonbold math in the Abstract.
% This preserves the distinction between vectors and scalars. However,
% if the conference you are submitting to favors bold math in the abstract,
% then you can use LaTeX's standard command \boldmath at the very start
% of the abstract to achieve this. Many IEEE journals/conferences frown on
% math in the abstract anyway.

%\IEEEpeerreviewmaketitle

%===============================================================================
\section{Introduction}

Increasing the complexity of systems proclaims the outsourced manufacturing concept nowadays. This raises a lot of trust issues in the design industry in many directions. Anyone with access to any step of the manufacturing process could alter the final product to inject a Hardware Trojan. Malicious circuitry can be injected by fabrication facilities or third party IP owners. This threatens design community as the fabrication process is obscured from designers and the details of third party IPs are hidden to protect IP owners' rights{\color{blue}~\cite{journal:Teh2010}}. 

Hardware Trojan appears to be one of the most important topics as the use of silicon chips in different applications becomes very popular, varying from cell phones, cars, to strategically important military devices. It is important to provide methods that resolve the trust issues between fabrication facilities, designers, and end-users. End-users need to make sure that products are not controlled by unknown entities, are stable enough, and will not leak critical information. Maintaining technology secrets of the fabrication facilities and design royalties of third party IP owners raises the difficulty of Hardware Trojan detection and protection. Homomorphic encryption may be used to solve this issue and defeat Hardware Trojans.

\begin{figure}[htb]
\centering
\includegraphics*[width = 0.45\textwidth]{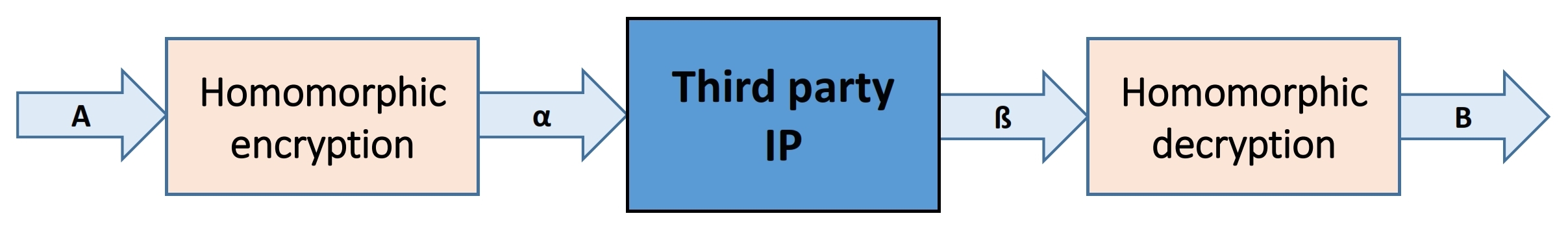}
\caption{Homomorphic encryption to protect from hardware Trojan.}
\label{fig:homop}
\end{figure}

In general, homomorphic encryption is a type of encryption, which allows specific types of operations to be carried out on ciphertext and generates an encrypted result which, when decrypted, matches the result of operations performed on the plaintext. This is a desirable feature that has been utilized in many modern systems{\color{blue}~\cite{conf:Rout12, conf:Hrestak14}}. 
%, conf:Zhao14}
In this work, we introduce the idea of using homomorphism to defeat Hardware Trojan injected in third party IPs. Consider the case where a third party IP is needed to carry out some operation on data $A$ and will produce output data $B$. Fig.~\ref{fig:homop} shows the ideal world, where the third party IP does not have any access to the real data as it is homomorphically encrypted. This will give us the capability to carry out the required operation by the third party IP without revealing the original data. Thus, we can retrieve the result $B$ after the decryption process. Full homomorphism (FH) evaluates any arbitrary depth circuit on ciphertexts, whereas partial homomorphism (PH) supports one type of operations only, addition or multiplication. 

%(Old)Hardware Trojan detection methods have been developed to ensure that chips are Trojan-free. These methods either try to detect the existence of a Trojan with high probability by studying side channels{\color{blue}~\cite{ conf:Jin2008, conf:Ban2008, conf:Pot2009, conf:Ban2009a, conf:Du2010, conf:Kou2010}}, or try to introduce architectural modifications to prevent embedding Tojan task{\color{blue}~\cite{conf:Sal2009, conf:Ban2009b, journal:Rad2010, conf:Raj2011}}. However, these methods mainly depend on comparing the suspected chip with a golden one (a known non-infected chip). In practice, a golden chip might not always be readily available, especially when using third party IPs. Attempts to depend on using the system integrator's design specifications for comparisons were introduced{\color{blue}~\cite{conf:zha2011}}.

The key contributions of this paper include: 
%\begin{enumerate}
\begin{compactenum}
\item Discussing new ideas to have a blind data processing by the third party IP with a minimum cost.
\item Implementing ElGamal encryption scheme, which is multiplicative homomorphic and the CRT-based Elgamal (CEG) encryption scheme, which is additive homomorphic, on a low-cost FPGA and showing the resource utilization, performance, and power analysis of both schemes.
\item Introducing a dual-circuit design that supports both, multiplicative and additive homomorphic properties and providing the obtained savings on area and power over a regular design that has no resource sharing. 
\end{compactenum}
%\end{enumerate}

The rest of the paper is organized as follows. Section~\ref{sec:relwrk} summarizes the related work. Overview about homomorphism and the utilized schemes are given in Section~\ref{sec:backgrnd}. Hardware Trojan protection using PH is introduced in Section~\ref{sec:prot}. Experimental evaluation and estimation of the overhead of the proposed methods are shown in Section~\ref{sec:expr}. Section~\ref{sec:conc} concludes the work.

%===============================================================================
\section{Related work} \label{sec:relwrk}

Recently, some architectural methodologies try to increase the chances of the activation of a Hardware Trojan during testing. Salmani \textit{et. al} increased the Trojan activity by inserting dummy flip-flops in the design{\color{blue}~\cite{conf:Sal2009}}. They chose the locations of the inserted flip-flops based on a transition probability threshold. Rajendran \textit{et al.} introduced a methodology for securing all the gates of the design using ring oscillators{\color{blue}~\cite{conf:Raj2011}}. They added extra logic that converts paths of the circuit into ring oscillators. Changes in the frequency of the ring oscillators were used to detect the presence of Trojans. Al-Anwar \textit{et al.} in{\color{blue}~\cite{conf:Spyware}} developed a novel method for the protection against a hardware Spyware that depends basically on decreasing the probability of seeking sensitive information. They introduced multiplexing between multiple variants implementation. Then, they used cyclic redundancy check (CRC) to detect the infected IP. In{\color{blue}~\cite{conf:Euro}}, Al-Anwar \textit{et al.} suggested obfuscating the output of the suspected IP before sending out data, then undoing that obfuscation at the input of the receiver in order to protect data from leaking and avoid injected triggering. They introduced using either RC4 or a simple obfuscating function. Ibn Ziad \textit{et al.} injects a Hardware Trojan in a voting machine to tamper voting results{\color{blue}~\cite{conf:voting}}. The attack depends mainly on the unused bits. They provided a protection technique against the proposed attack and showed its overhead. Al-Anwar \textit{et al.} in{\color{blue}~\cite{conf:canada}} introduced multiplexing reconfigurable IPs' outputs and CRC Trojan detection scheme (MCRC) method in order to decrease the probability of leaking critical information by a hacked IP. They also suggested using partial reconfiguration technology to remove an infected IP. %Furthermore, they introduced voting between the output of odd number (at least three) of the same IP from different vendors to get a safe output.

%Banga and Hsiao used voltage inversion at alternating levels of the circuit to increase the power consumption of an infected circuit{\color{blue}~\cite{conf:Ban2009b}}.

%TODO rewrite the following 3 paragraphs (Done by Tarek) 
Side-channel dependent methodologies for Hardware Trojan detection aim to localize the impact of the Trojan on the circuit without activating it. Their main idea is to try to detect the presence of a Trojan with high probability via detecting the overload of the Trojan circuit on different circuit parameters; such as the delay or the power as compared to a non-infected circuit. Rad \textit{et al.} studied the impact of a Trojan on the power supply transient current of an IC using statistical methods{\color{blue}~\cite{journal:Rad2010}}. Jin and Makris used path delay analysis to detect Trojans{\color{blue}~\cite{conf:Jin2008}}. Moreover, gate-level characterization techniques accompanied by statistical methods were used to detect Hardware Trojans{\color{blue}~\cite{journal:Wei2011}}. 

%Banga \textit{et al.} proposed a test vector generation method, which can be used to differentiate between the side-channel waveforms of a Trojan infected and a non-infected circuit{\color{blue}~\cite{conf:Ban2009a}}.
%Moreover, Gate-level characterization techniques accompanied by statistical methods were used to detect Hardware Trojans. These methods depend mainly on gate-level delay or power characterization, or both{\color{blue}~\cite{conf:Alk2009, conf:Pot2009, conf:Kou2010, journal:Wei2011}}. 

%Wang \textit{et. al} utilized localized current analysis for Trojan detection. They measured power from multiple ports to detect the impact of the Trojan on power{\color{blue}~\cite{conf:Wang2008b}}.

%In general, regular testing methods failed to detect the faults caused by Hardware Trojans because Trojans are not expected to be activated during testing. This reason is Trojan circuits are usually designed to be activated using a rare trigger. Recently, various methods have been specifically designed with Trojan detection in mind. One can classify these Hardware Trojan detection methodologies into: (1) side-channel dependent methodologies and (2) architectural methodologies{\color{blue}~\cite{journal:Teh2010}}. 

Unfortunately, all Hardware Trojan detection methods require the presence of a non-infected (golden) chip. That requirement represents a real problem as it is feasible only if the design does not contain third party IPs{\color{blue}~\cite{journal:Teh2011}}. But, if the system designer integrates third party IPs in the design, these methods become less practical. Zhang and Tehranipoor tried to provide an alternative to using a golden design by using code coverage analysis, formal verification, and ATPG methods to achieve high confidence in whether the circuit is Trojan-free or Trojan-inserted{\color{blue}~\cite{conf:zha2011}}. Baumgarten \textit{et al.} suggested using reconfigurable logic barriers within a design to prevent the activation and operation of Hardware Trojans inserted during the manufacturing stage of an IC{\color{blue}~\cite{journal:Bau2010}}. %Waksman and Sethumadhavan presented a method that attempts to prevent Trojan triggering{\color{blue}~\cite{conf:Wak2011}}. Beaumont \textit{et al.} ran replica of a program on multiple processing elements in order to achieve protection form hardware Trojans{\color{blue}~\cite{conf:Bea2012}}.  
%===============================================================================
%\section{Partial Homomorphism Background} \label{sec:backgrnd}
\section{Background} \label{sec:backgrnd}

The aim of this section is to give a brief description about the idea of homomorphism, survey existing partial homomorphic encryption schemes, and discuss ElGamal security scheme.

\subsection{Partial Homomorphism (PH)} \label{sub:Phomo}
%From Practical HE A survey (need to be modified)
PH has been known for many years. It offers the ability to perform a certain type of operations, addition or multiplication, on ciphertexts without revealing data. For example, let us consider the two messages, $m_1$ and $m_2$, where both messages are encrypted and their ciphertexts are given by $E(m_1)$ and $E(m_2)$, respectively. If the multiplication of the two ciphertexts is equivalent to the ciphertext of the multiplication of the two messages as shown in (\ref{eqn:mul}), we call this a multiplicative homomorphic scheme. On the other hand, if the multiplication of the two ciphertexts equals the ciphertext of the addition of the two messages as shown in (\ref{eqn:add}), we call this an additive homomorphic scheme. 

\begin{equation}
E(m_1) \times E(m_2) = E(m_1 \times m_2)
\label{eqn:mul}
\end{equation} 

\begin{equation}
E(m_1) \times E(m_2) = E(m_1 + m_2)
\label{eqn:add}
\end{equation} 

%\begin{equation}
%E(m_1) = {m_1}^e, \; E(m_2) = {m_2}^e 
%\label{eqn:homo1}
%\end{equation}

%\noindent The multiplication of the two ciphertexts is equivalent to the ciphertext of the multiplication of the two messages as shown in (\ref{eqn:homo2}). 

%\begin{equation}
%E(m_1) \times E(m_2) = {(m_1 \times m_2)}^e = E(m_1 \times m_2)
%\label{eqn:homo2}
%\end{equation} 

%This is called multiplicative homomorphism. Other schemes allow implements the same idea but using addition, so they are called additive homomorphic schemes. Partial homomorphic schemes support only one type of operation on encrypted data.

%It is worth mentioning that PH is different from FH, which allows the efficient evaluation of an arbitrary depth circuit (composed of additions and multiplications) to be evaluated directly on ciphertexts. The first full homomorphic encryption scheme was introduced by Gentry{\color{blue}~\cite{phD:Gentry}} in 2009. However, current homomorphic encryption schemes are still not efficient enough for real time applications due to its very large ciphertext and public key sizes{\color{blue}~\cite{homoSurvey}}. Thus, we focus on PH in this paper. 

It is worth mentioning that PH is different from FH, which allows the efficient evaluation of an arbitrary depth circuit (composed of additions and multiplications) to be evaluated directly on ciphertexts. The first full homomorphic encryption (FHE) scheme was introduced by Gentry{\color{blue}~\cite{phD:Gentry}} in 2009. Since then, there has been some work done toward obtaining efficient hardware implementations of FHE schemes. Hardware building blocks for the lattice-based cryptosystem were considered by G\"{o}ttert \textit{et al.}{\color{blue}~\cite{Gottert:2012}}. Also, P\"{o}ppelmann and G\"{u}neysu introduced an efficient hardware implementation of ring-learning-with-errors (RLWE) based encryption{\color{blue}~\cite{Thomas12}}. However, these schemes are not practical for this application due to its very large ciphertext and public key sizes. Thus, we focus on PH in this paper. 

One of the earliest discoveries in the context of PH is the Goldwasser-Micali cryptosystem{\color{blue}~\cite{jour:Goldwasser84}}, whose security is based on the quadratic residuosity problem. It allows homomorphic evaluation of a bitwise exclusive-or. Other additive homomorphic encryption schemes that provide semantic security are Benaloh{\color{blue}~\cite{jour:Clarkson94}} and Paillier{\color{blue}~\cite{Paillier99}}. On the other hand, there exist two well-known schemes, which are multiplicative homomorphic schemes. The first one is the Rivest-Shamir-Adleman (RSA){\color{blue}~\cite{conf:RSA78}}, which is one of the most widely used public-key cryptosystems. The second is ElGamal encryption scheme{\color{blue}~\cite{ElGamal85}}, which is the selected cryptosystem to be used in our work.

\subsection{ElGamal Scheme} \label{sub:ElGamal}
ElGamal public-key cryptography algorithm is considered to be one of the efficient and popular algorithms that provides a high level of security. To illustrate its functionality, let us consider that a user called \textit{Alice} wants to send a private message $m$ to another user \textit{Bob}. ElGamal process works as follows. \textit{Bob} generates his keys. He chooses a secret random exponent $k$ and a generator $g$. So, his public key is $(g,h)$ where ${h = g^k (mod \: n)}$ and $n$ is a large prime. \textit{Alice} has to encrypt the message $m$ before sending it to \textit{Bob}. She generates a random exponent $l$ and sends the ordered pair $(C_1,C_2)$ to \textit{Bob}, where $C_1$ and $C_2$ are defined as~(\ref{eqn:C12}).    

\begin{equation}
C_1 = g^l (mod \: n), C_2 = h^l \times m (mod \: n)
\label{eqn:C12}
\end{equation}

%in (\ref{eqn:C1}) and (\ref{eqn:C2}), respectively.   

%\begin{equation}
%C_1 = g^l (mod \: n)
%\label{eqn:C1}
%\end{equation}

%\begin{equation}
%C_2 = h^l \times m (mod \: n)
%\label{eqn:C2}
%\end{equation}

\noindent \textit{Bob} can easily decrypt the ciphertext using (\ref{eqn:m1}).

\begin{equation}
m = {C_1}^{-k} \times C_2 (mod \: n)
\label{eqn:m1}
\end{equation}

\noindent This encryption scheme is homomorphic with respect to multiplication as if $(x_1,y_1)$ and $(x_2,y_2)$ are valid encryptions for messages $m_1$ and $m_2$, with the same key, then $(x_1x_2,y_1y_2)$ is a valid encryption of $m_1m_2$. Hu \textit{et al.} proposed a simple modification to make ElGamal additively homomorphic by placing the message $m$ in the exponent{\color{blue}~\cite{homoCRT}}. So, if we encrypt two messages $m_1$ and $m_2$ using (\ref{eqn:C12}) but multiply $h^l$ with $g^m$ instead of $m$, the multiplication of the two ciphertexts results in a valid encryption of $g^{m_1+m_2}$. The problem here is that recovering the message involves solving a discrete logarithm problem (DLP) and this is precisely the problem whose difficulty ensures security. To solve this problem, they introduced a new scheme, called CRT-based ElGamal Scheme (CEG), which uses the Chinese Remainder Theorem (CRT) to replace one DLP in a large space by several similar problems in a more tractable search space. This allows for easily obtaining $m_1 + m_2$, while retaining the full security of the scheme, as shown later.
 
\subsection{CRT-based ElGamal (CEG) Scheme}
To illustrate how CEG works, let us reuse the previous example of \textit{Alice} and \textit{Bob}. In the first step, \textit{Bob} also chooses a secret random exponent $k$ and a generator $g$. He also chooses $d_i$ for $i = 1,\dots,t$ such that $gcd(d_i,d_j)=1$ for $i \neq j$. So, his public key is $(g,h,(d_1,\dots,d_t))$, where ${h = g^k (mod \: n)}$ and $n$ is a large prime. For encryption, \textit{Alice} sends the encryption of message $m$ as a t-tuple of pairs $(C_1,C_2)$ by using (\ref{eqn:CEGC12}).

\begin{equation}
C_1 = g^{l_i} (mod \: n), C_2 = h^{l_i} \times g^{m_i} (mod \: n) 
\label{eqn:CEGC12}
\end{equation}

%\begin{equation}
%C_1 = g^{l_i} (mod \: n), i = 1,\dots,t 
%\label{eqn:CEGC1}
%\end{equation}

%\begin{equation}
%C_2 = h^{l_i} \times g^{m_i} (mod \: n), i = 1,\dots,t
%\label{eqn:CEGC2}
%\end{equation}

\noindent where $m_i = m \; (mod \; d_i)$ and $l_i$ is a generated random exponent for $i = 1,\dots,t$. \textit{Bob} can decrypt the ciphertext using (\ref{eqn:mCEG1}) and (\ref{eqn:mCEG2}).

\begin{equation}
m = CRT^{-1} [(log_g(C_{2_i} \times C_{1_i}^{-k} (mod \: n)) , i = 1,\dots,t)]
\label{eqn:mCEG1}
\end{equation}

\begin{equation}
CRT^{-1}[C_i] = \sum_{i = 1}^{t}C_i \: \frac{d}{d_i} (\frac{d}{d_i}^{-1} mod \; d_i) \: mod \; d
\label{eqn:mCEG2}
\end{equation}

%\noindent where $CRT^{-1}[C_i] = \sum_{i = 1}^{t}C_i \: \frac{d}{d_i} (\frac{d}{d_i}^{-1} mod \; d_i) \: mod \; d$ for $i = 1,\dots,t$. Correctness and efficiency of the illustrated scheme is discussed in details in{\color{blue}~\cite{homoCRT}}. As a part of this work, we implement the CEG scheme in hardware and show its resource utilization and power consumption.

\noindent Correctness and efficiency of the illustrated scheme is discussed in details in{\color{blue}~\cite{homoCRT}}. As a part of this work, we implement the CEG scheme in hardware and show its resource utilization and power consumption.
%===============================================================================
\section{Hardware Trojan protection using PH} \label{sec:prot}

Here, we introduce our suggested methods for defeating Hardware Trojan in third party IPs. First, we propose two schemes that support PH for the third party IP, which performs one type of operation (multiplication only or addition only). Then, we combine the two methods in a dual-circuit design that supports both multiplication and addition to satisfy applications that utilize the two operations.  

\subsection{Sufficient PH Support} 

%Upon classifying the IPs based on processing type, one concludes that there is no need to afford the high cost of FH if the third party IP does only one type of operation. It is totally sufficient to have PH encryption/decryption before/after the suspected IP. In other words, if the suspected IP does addition operation, it is enough to support one of the additively homomorphic schemes mentioned before in Subsection~\ref{sub:Phomo}. For non computational suspected IPs, it is adequate to do simple obfuscation functions before the suspected IP and do the inverse of that function afterwords. In{\color{blue}~\cite{conf:Wak2011}}, Waksman and Sethumadhavan suggested to xor input data before and after the suspected IP. We discuss two partial homomorphic hardware implementations based on ElGamal encryption scheme described in Subsection~\ref{sub:ElGamal}. The first implementation is the main ElGamal encryption/decryption scheme{\color{blue}~\cite{ElGamal85}}, which is a multiplicative homomorphic scheme. The second one is the CEG scheme{\color{blue}~\cite{homoCRT}}, which is an additive homomorphic scheme.  

Upon classifying the IPs based on processing type, one concludes that there is no need to afford the high cost of FH if the third party IP does only one type of operation. It is totally sufficient to have PH encryption/decryption before/after the suspected IP. In other words, if the suspected IP is used in an electronic voting system and only does addition operation to count votes on the server side{\color{blue}~\cite{conf:voting}}, it is enough to support one of the additively homomorphic schemes mentioned before in Subsection~\ref{sub:Phomo}. For non computational suspected IPs, it is adequate to do simple obfuscation functions before the suspected IP and do the inverse of that function afterwords. Here, we discuss two partial homomorphic hardware implementations based on ElGamal encryption scheme described in Subsection~\ref{sub:ElGamal}. The first implementation is the main ElGamal encryption/decryption scheme{\color{blue}~\cite{ElGamal85}}, which is a multiplicative homomorphic scheme. The second one is the CEG scheme{\color{blue}~\cite{homoCRT}}, which is an additive homomorphic scheme.  

\subsubsection{\textbf{Elgamal Scheme Implementation}}

\begin{figure}
    \centering
    \begin{subfigure}[htb]{0.73\linewidth}%hbt
        \centering
            \includegraphics[width = \textwidth]{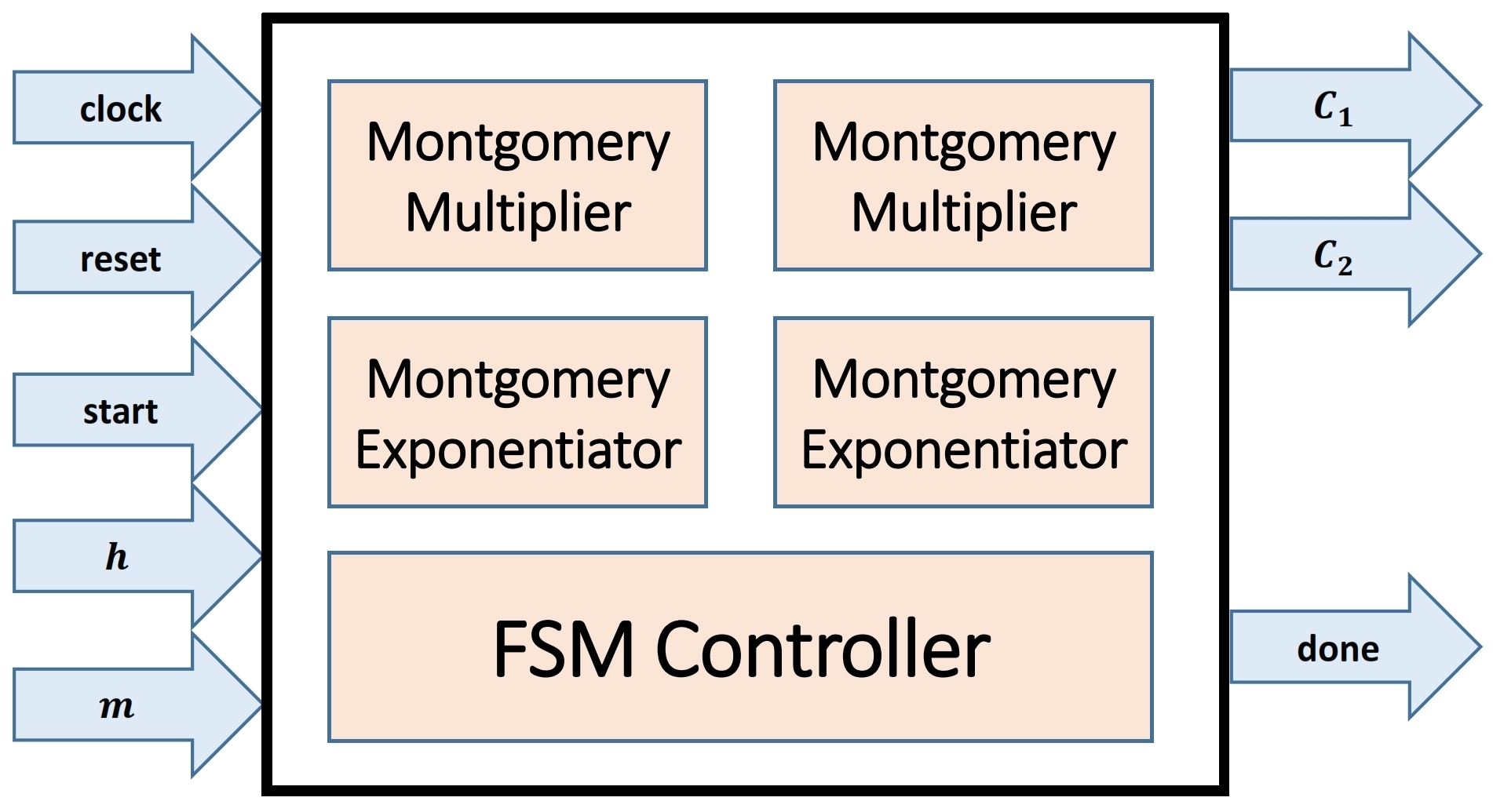}
            \caption{Encryption.}
    \end{subfigure}
    \begin{subfigure}[htb]{0.73\linewidth}%hbt
        \centering
            \includegraphics[width = \textwidth]{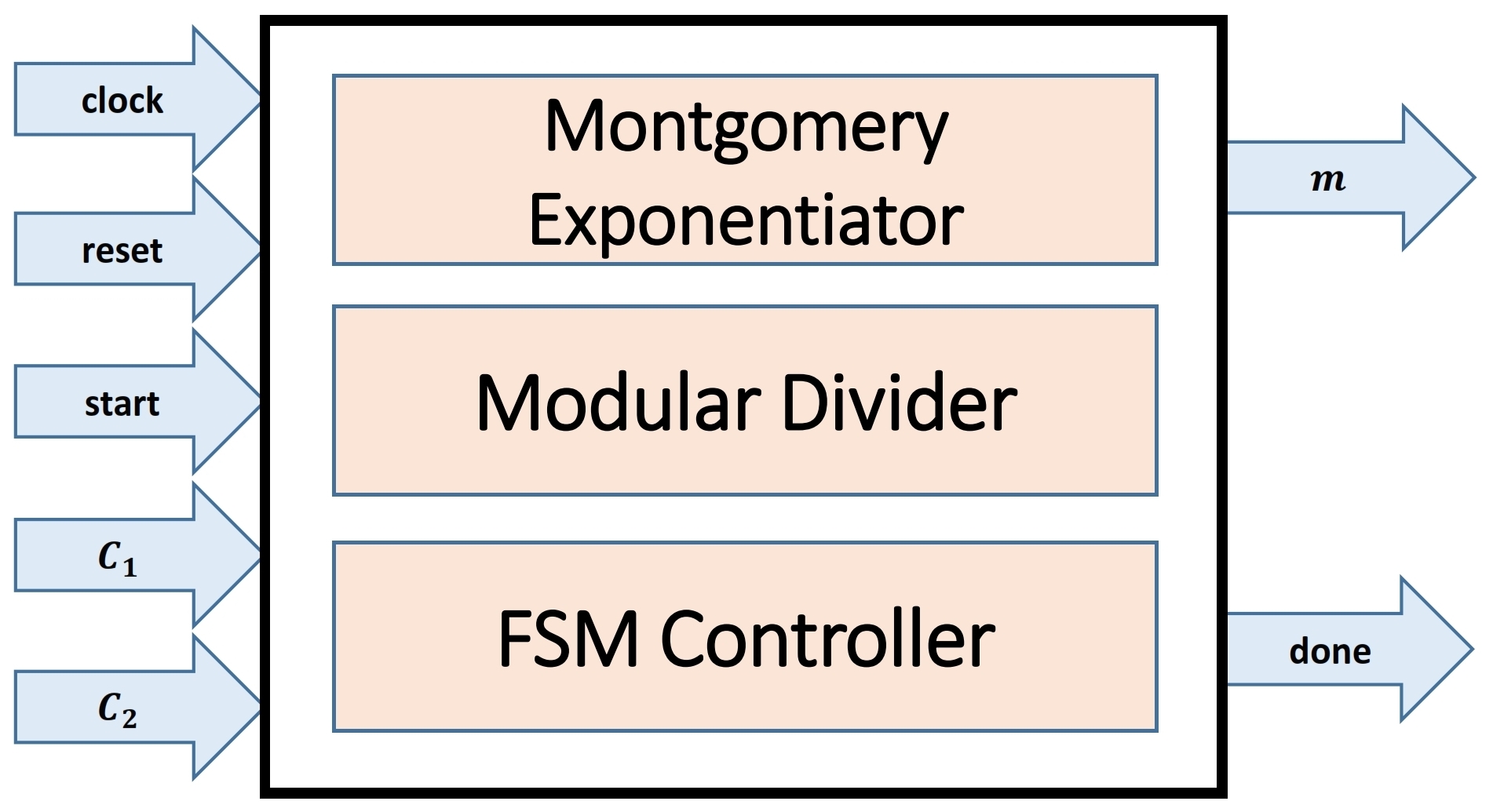}
            \caption{Decryption.}
    \end{subfigure}
    \caption{Block diagram for ElGamal encryption/decryption scheme.}
    \label{fig:Mul}
\end{figure}

Fig.~\ref{fig:Mul} shows the block diagram for our implementation of ElGamal encryption/decryption scheme. The encryption module consists of two Montgomery modular multipliers, two Montgomery modular exponentiators, and a finite state machine (FSM) controller that is responsible for synchronizing other components' inputs and outputs to perform the encryption operations defined in (\ref{eqn:C12}). The decryption module consists of one Montgomery modular exponentiator, one modular divider, and a FSM controller that is also responsible for synchronizing other components' inputs and outputs to perform the decryption operations defined in (\ref{eqn:m1}). Both modules use a clock and reset signals as inputs. Reset and done signals are utilized to indicate the start and the end of module operations. The message $m$, ciphertexts $C_1 \; and \; C_2$, and the public key $h$ are all $k$ bits vectors, where $k$ is a user-defined integer.   

Montgomery multipliers were used in the design as the Montgomery's algorithm{\color{blue}~\cite{Montgomery85}} is the most widely used algorithm for efficient modular multiplication. Other multiplication methods like the \textit{multiply and reduce} and \textit{double, add, and reduce} are computationally more complex{\color{blue}~\cite{book:Deschamps09}}. The binary Montgomery multiplier employs only simple addition, subtraction, and shift operation to avoid trial division, which is a critical and time-consuming operation in conventional modular multiplication. In fact, this multiplier computes $Z = X \times Y \times R^{-1} mod \; M$ instead of $Z = X \times Y mod \; M$, where $R$ is a chosen integer that should be a power of two and relatively prime to $M$. So, in this case, the operands need to be converted into and out of Montgomery's domain each time this multiplier is used. 

In general, modular exponentiation is usually accomplished by performing repeated modular multiplications. For our modular exponentiators, the LSB-first algorithm using Montgomery multiplication is used. This algorithm computes $Z = Y^X mod \; M$ in $k$ executions of a loop that, in turn, includes at most two Montgomery multiplication operations, which are executed concurrently. That improves the performance of the module{\color{blue}~\cite{book:Deschamps09}}. 

The decryption part of the scheme includes the usage of a modular divider module. We implemented the plus-minus algorithm as it gives the shortest computation time with a cost-effective area{\color{blue}~\cite{jour:Div06}}. The key generation module consists mainly of a Montgomery exponentiation circuit and a true random number generator (TRNG) module, which is not in the scope of this paper. Finally, it is worth noting that the usage of only one multiplier and one exponentiator is enough to achieve the desired encryption results, but that results in a high critical path delay. 
 
\subsubsection{\textbf{CEG Scheme Implementation}}

\begin{figure}
    \centering
    \begin{subfigure}[bh]{0.73\linewidth}%hbt
        \centering
            \includegraphics[width = \textwidth]{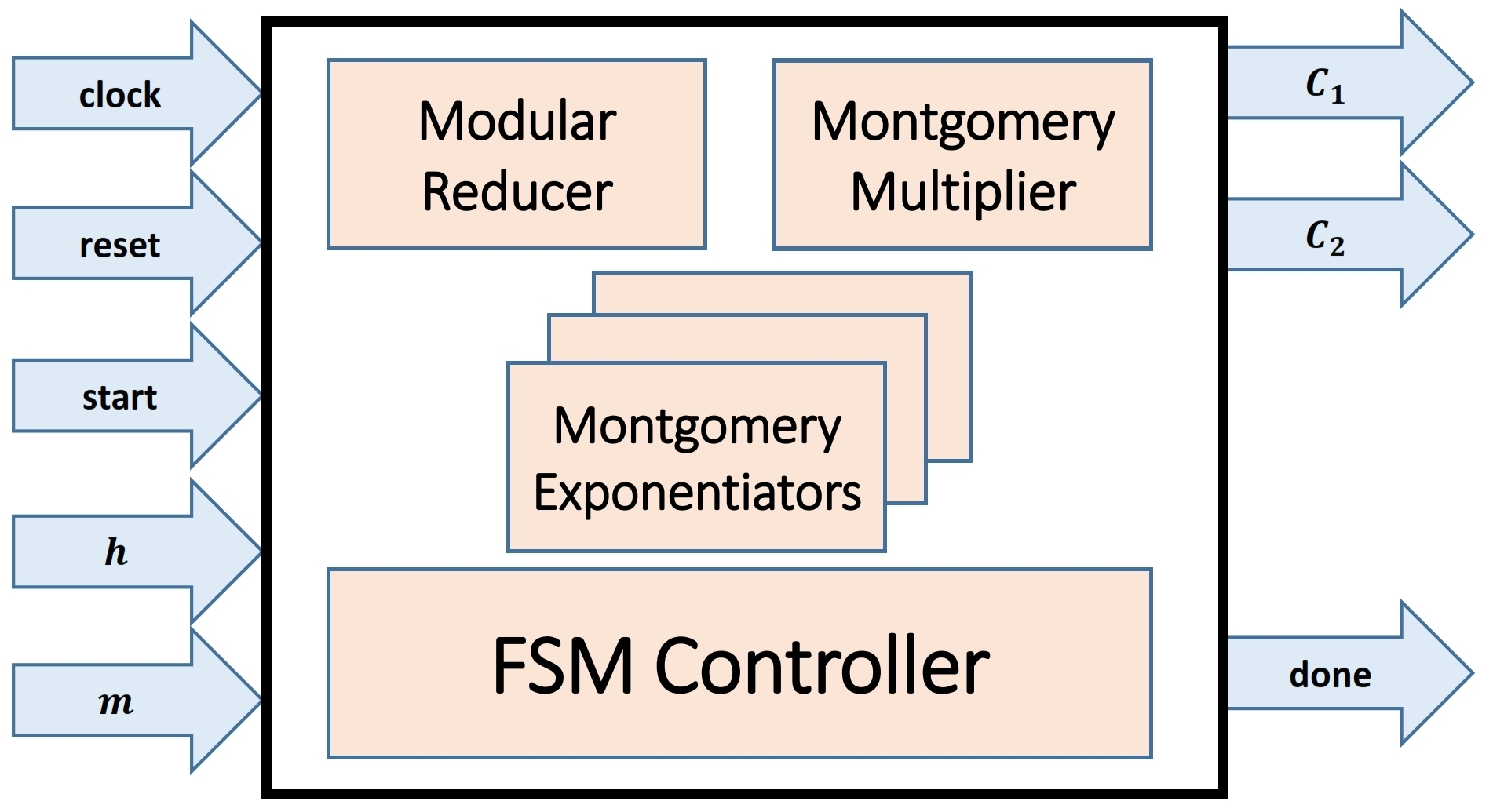}
            \caption{Encryption.}
    \end{subfigure}
    \begin{subfigure}[bh]{0.73\linewidth}%hbt
        \centering
            \includegraphics[width = \textwidth]{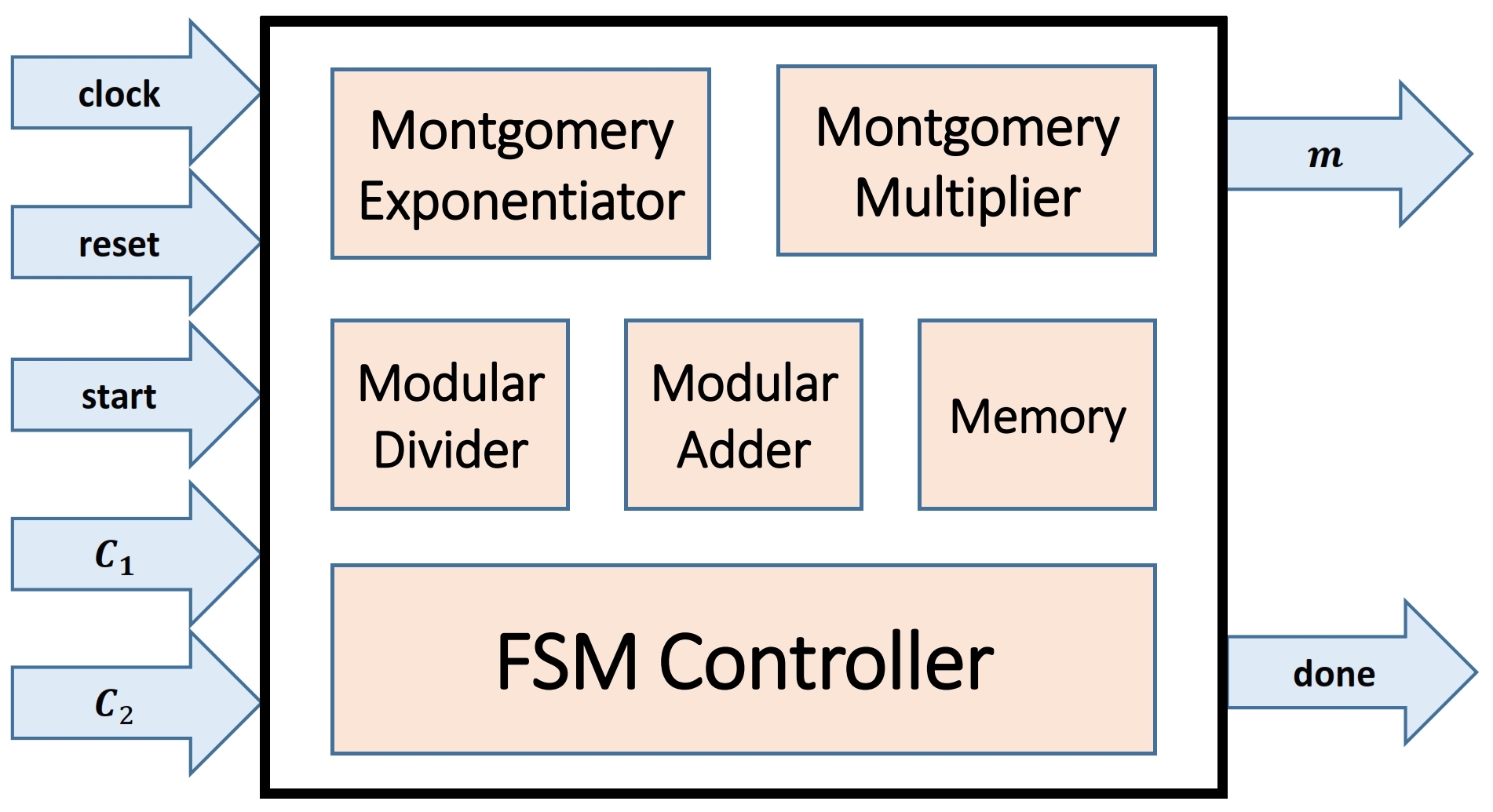}
            \caption{Decryption.}
    \end{subfigure}
    \caption{Block diagram for the CRT-based ElGamal (CEG) encryption/decryption scheme.}
    \label{fig:Add}
\end{figure}

Fig.~\ref{fig:Add} shows the block diagram for our implementation of the CEG encryption/decryption scheme. This design is quietly different from ElGamal design discussed before as the encryption operations defined in (\ref{eqn:CEGC12}) requires the usage of multiple Montgomery exponentiators. As the timing delay needed by one exponentiator is more than the delay of a single multiplier, the FSM controller is modified to utilize only one Montgomery multiplier. A modular reducer circuit is used to handle the operation of reducing $m$ into several $m_i$ based on the relation of $m_i = m (mod \: d_i)$ for $i = 1,\dots,t$.  

For the decryption module, it consists of one Montgomery modular exponentiator, one Montgomery modular multiplier, one modular divider, one modular adder, FSM controller, and a single block of memory used to facilitate the implementation of the inverse CRT needed in (\ref{eqn:mCEG1}){\color{blue}~\cite{homoCRT}}. Input and output vectors are now $k \times t \; bits$ instead of $k \; bits$, where $t$ is the number of ciphertext pairs.

\subsection{Dual-Circuit Design} 

The main motivation for this design is that some third party IPs require the usage of more than one single type of operation. For instance, an IP may need to perform both addition and multiplication but not at the same time. One can imagine the functionality of that IP as a simple ALU that uses a selection line to switch its mode between two different operations. In this case, using one type of partial homomorphic schemes would not be sufficient. We have to implement two different schemes, such as implementing the two schemes described above, in order to prevent the attacker from revealing the ALU input and output data. We suggest a solution for this issue by combining the two previously proposed schemes, ElGamal and the CEG, in a single dual-circuit design. Thus, the proposed design supports both additive and multiplicative homomorphism.   

%Fortunately, we do not have to implement all possible functions to achieve full homomorphism. Addition and multiplication on an any non-trivial ring are enough as the sum and product is equivalent to XOR and AND gates, respectively. If one can compute sums and products on encrypted bits, he can compute any function on encrypted inputs{\color{blue}~\cite{project:wade}}. Thus, we suggest to combine the two previously proposed schemes, ElGamal and the CEG, to implement a full homomorphic ElGamal scheme. 

Furthermore, we try to share resources as much as we can between the two schemes in order to have minimal design cost. For example, computing $C_1$ in (\ref{eqn:C12}) and (\ref{eqn:CEGC12}) needs an exponentiation operation. The same situation occurs when computing $C_2$ as we need an exponentiation operation followed by a multiplication operation. The only difference is that the modified versions in (\ref{eqn:CEGC12}) reuse their modules many times based on the value of $t$. Thus, we use the duality concept that enables us of sharing as much resources between the two circuits in order to reduce the design area. As the CEG scheme uses the same basic blocks of ElGamal scheme with some additional blocks, we depend on the same architecture shown in Fig.~\ref{fig:Add} and add a $select$ signal that chooses between the multiplicative homomorphic one and the additive homomorphic algorithms. The FSM controller is modified to be able of handling the two cases with the same building modules. The case is the same for the key generation and decryption modules. 

By using this simple idea, we manage to decrease the area cost a lot and allow for the two homomorphic properties to be available on a single module. That completely solves the issue of the third party IP, which needs to perform both addition and multiplication operation. Moreover, another possible example for an application that needs the availability of both homomorphic operations is when we have two unique IPs in a design and the first IP performs addition while the second IP performs multiplication. Assuming that both IPs will not work on the same time, one can instantiate only one instance of our dual-circuit module and control its functionality to perform the needed operation of any of the two IPs, when needed, with the minimal cost in area and power consumption.  

%===============================================================================
\section{Experimental results} \label{sec:expr}

This section evaluates the performance of our proposed methods, described in Section~\ref{sec:prot}, in terms of resource utilization, delay, and power consumption. The proposed methods are implemented on Xilinx Spartan-6 XC6SLX75 with FGG484 package and -2 speed grade. The area and performance results are obtained from the Xilinx ISE 14.6 tool after place and route analysis. The power is calculated using Xilinx Xpower Analyzer with 100 MHz clock.   

%{\color{blue}~\cite{SheetSpartan}}
%{\color{blue}~\cite{IncXilinx}}
\subsection{PH Schemes Results}
Table \ref{tab:PartialRes} shows the top-level module resource utilization of our two partial homomorphic encryption/decryption schemes, ElGamal and CEG, using vectors of size equals 8 bits. %The table also gives the area overhead of the CRT-based idea that turns ElGamal scheme from being multiplicative homomorphic to be additive homomorphic. Although the overhead seems to be high, the idea is appealing as it offers the opportunity to build a full homomorphic scheme based on a partial homomorphic one. 

\begin{table}[b]
  \centering
  \caption{Resource utilization of ElGamal and CRT-based ElGamal (CEG) encryption/decryption schemes for k = 8 bits.}
    \tabcolsep=0.11cm 
    %\resizebox{0.48\textwidth}{!}{
    \begin{tabular}{|l||c|c||c|c|}
    \toprule
    & \multicolumn{2}{c}{Encryption }  & \multicolumn{2}{c}{Decryption} \vline\\
    & ElGamal  & CEG   & ElGamal  & CEG   \\
    \midrule
    Number of Registers & 295   & 614   & 207   & 364   \\
    Number of LUTs      & 420   & 715   & 259   & 442   \\
    Number of BRAMs     & 0     & 0     & 0     & 1     \\
    \bottomrule
    \end{tabular}%}%
  \label{tab:PartialRes}%
\end{table}%

%\noindent The overhead column is calculated using (\ref{eqn:AreaOver}).

%\begin{equation}
%Overhead (\%) = \cfrac{CEG \; area - ElGamal \; area}{ElGamal\; area} \times 100.
%\label{eqn:AreaOver}
%\end{equation}

Table \ref{tab:PartialTiming} shows the maximum operating frequency of the two proposed partial homomorphic schemes along with the needed number of cycles to finish their work. 

\begin{table}[b]
  \centering
  \caption{Timing performance of ElGamal and CRT-based ElGamal (CEG) encryption/decryption schemes for k = 8 bits.}
    \tabcolsep=0.11cm 
    %\resizebox{0.48\textwidth}{!}{
    \begin{tabular}{|l||c|c||c|c|}
    \toprule
                  & \multicolumn{2}{c}{Encryption } & \multicolumn{2}{c}{Decryption} \vline\\
                  & ElGamal & CEG     & ElGamal & CEG \\
    \midrule
    Frequency (MHz)& 161.277 & 164.352 & 123.870 & 121.862 \\
    No. of Cycles & 171   & 480   & 153   & 512 \\
 %   Time (us) & 1.06029 & 2.92056 & 1.23517 & 4.20147\\
    \bottomrule 
    \end{tabular}%}%
  \label{tab:PartialTiming}%
\end{table}%

From power prospective, Table \ref{tab:PartialPower} shows the power analysis for ElGamal encryption/decryption scheme and the CRT-based one. It was found that the dynamic power slightly decreased in case of encryption and increased in case of decryption due to the usage of the memory component and its logic controller in decryption. The leakage power remains constant in the both cases.

\begin{table}[t]
  \centering
  \caption{Power consumption (mW) of ElGamal and CEG encryption/decryption schemes for k = 8 bits.}
    \tabcolsep=0.11cm 
    %\resizebox{0.4\textwidth}{!}{
    \begin{tabular}{|l||c|c||c|c|}
    \toprule
                  & \multicolumn{2}{c}{Encryption } & \multicolumn{2}{c}{Decryption} \vline\\
                  & ElGamal & CEG     & ElGamal & CEG \\
    \midrule
    Clocks  & 5.65 & 7.87 & 4.21 & 5.87 \\
    Logic   & 3.84 & 5.47 & 2.70 & 3.69 \\
    Signals & 2.82 & 4.69 & 2.01 & 3.23 \\
    BRAMs   & 0.00 & 0.00 & 0.00 & 0.74 \\
    IOs     & 16.51 & 8.99 & 5.23 & 2.74 \\
    Leakage & 65.00 & 65.00 & 64.00 & 64.00 \\
    \midrule
    Total   & 93.82 & 92.02 & 78.15 & 80.27 \\
    \bottomrule 
    \end{tabular}%}%
  \label{tab:PartialPower}%
\end{table}%

\subsection{Dual-Circuit Design Results}

Here, we compare the results of our proposed dual-circuit design to using regular two IPs, one for ElGamal and another for CEG design without any resource sharing between them. We want to address the effect of our resource sharing. In order to differentiate between the two designs, we call the first design, \textit{Dual ElGamal}, while the second design is called \textit{Regular ElGamal}.    

Firstly, Table \ref{tab:FullRes} shows the area reduction that results from using our \textit{Dual ElGamal} design over \textit{Regular ElGamal} design. The area reduction column is calculated using (\ref{eqn:AreaRed}). It is clear that the idea of dual-circuit design has greatly improved the usage of hardware resources.

%Using any of the full homomorphic schemes mentioned in the literature will not be fair to evaluate our dual ElGamal design as they are not based on the same security algorithm. To handle this issue, we compare the results of our proposed design to a full homomorphic (FH) ElGamal design that combines the two partial homomorphic schemes, ElGamal and CEG. Firstly, table \ref{tab:FullRes} shows the amount of area reduction that results from using our dual ElGamal design over the FH ElGamal design. The area reduction column is calculated using (\ref{eqn:AreaRed}). It is clear that the idea of dual-circuit design has greatly improved the usage of hardware resources.

\begin{equation}
Reduction (\%) = \cfrac{Regular \; area - Dual \; area}{Regular\; area} \times 100.
\label{eqn:AreaRed}
\end{equation}
 
\begin{table}[b]
  \centering
  \caption{Area reduction of our Dual ElGamal design over the Regular ElGamal design for k = 8 bits.}
    \tabcolsep=0.11cm 
    %\resizebox{0.48\textwidth}{!}{
    \begin{tabular}{|l||c|c|c||c|c|c|}
    \toprule
                    & \multicolumn{3}{c}{Encryption}  & \multicolumn{3}{c}{Decryption} \vline \\
    & Regular & Dual    & Area      & Regular & Dual    & Area      \\
    & ElGamal & ElGamal & reduction & ElGamal & ElGamal & reduction \\
    &         &         &   (\%)    &         &         &   (\%) \\
    \midrule
    Registers & 909   & 635   & 30.14 & 536   & 364   & 32.09 \\
    LUTs      & 1137  & 735   & 35.36 & 626   & 457   & 26.99 \\
    BRAMs     & 0     & 0     & 00.00 & 1     & 1     & 00.00 \\
    \bottomrule 
    \end{tabular}%}%
  \label{tab:FullRes}%
\end{table}%

Table \ref{tab:FullTiming} gives the maximum operating frequency of our \textit{Dual ElGamal} design and the \textit{Regular ElGamal} design using vectors of size $k = 8 \; bits$. The number of cycles here represents the clock cycles needed to perform one multiplicative homomorphic operation followed by one additive homomorphic operation. The needed number of cycles to get the final output is the same in both designs, except that the encryption part of our dual designs utilizes more clock cycles. That is due to the usage of only one Montgomery multiplier instead of two, as illustrated in Section~\ref{sec:prot}. 

\begin{table}[t]
  \centering
  \caption{Timing comparisons between our Dual ElGamal design and the Regular ElGamal design for k = 8 bits.}
    \tabcolsep=0.11cm 
    %\resizebox{0.48\textwidth}{!}{
     \begin{tabular}{|l||c|c||c|c|}
    \toprule
                    & \multicolumn{2}{c}{Encryption }  & \multicolumn{2}{c}{Decryption} \vline \\
                    & Regular & Dual    & Regular & Dual \\
                    %& ElGamal & ElGamal & ElGamal & ElGamal \\
    \midrule
    Frequency (MHz) & 161.277 & 158.51  & 117.099 & 121.344 \\
    No. of Cycles   & 651     & 662     & 665     & 665 \\
   % Time (us)       & 4.03653 & 4.17639 & 5.67896 & 5.48029 \\
    \bottomrule
    \end{tabular}%}%
  \label{tab:FullTiming}%
\end{table}%

From power prospective, Table \ref{tab:FullPower} shows the power analysis for our \textit{Dual ElGamal} design and the \textit{Regular ElGamal} design. The usage of the duality idea results in an obvious improvement in total power consumption as it eliminates the power consumed by the duplicated modules. The savings in power consumption are 20.44\% for encryption and 12.26\% for decryption. 

\begin{table}[t]
  \centering
  \caption{Power consumption (mW) of our Dual ElGamal design and the Regular ElGamal design for k = 8 bits.}
    \tabcolsep=0.11cm 
    %\resizebox{0.4\textwidth}{!}{
    \begin{tabular}{|l||c|c||c|c|}
    \toprule
                    & \multicolumn{2}{c}{Encryption }  & \multicolumn{2}{c}{Decryption} \vline \\
                    & Regular & Dual    & Regular & Dual \\
                    %& ElGamal & ElGamal & ElGamal & ElGamal \\
    \midrule
	 Clocks  & 11.78 & 6.89 & 8.78 & 4.86 \\
    Logic   & 9.25 & 6.29 & 5.91 & 3.82 \\
    Signals & 8.14 & 6.02 & 5.67 & 3.49 \\
    BRAMs   & 0.00 & 0.00 & 0.74 & 0.74 \\
    IOs     & 25.27 & 10.83 & 5.67 & 3.61 \\
    Leakage & 65.00 & 65.00 & 65.00 & 64.00 \\
    \midrule
    Total   & 119.44 & 95.03 & 91.77 & 80.52 \\
    \bottomrule 
    \end{tabular}%}%
  \label{tab:FullPower}%
\end{table}%

%We may add one figure that gives a resource utilization estimation in case of using higher vectors sizes. 

%===============================================================================
\section{Conclusion} \label{sec:conc}

In this work, we highlighted the importance of homomorphic encryption in defeating Hardware Trojans in third party IPs. As PH is sufficient enough with some third party IPs, we implemented two designs that supports PH (multiplicative only and additive only) based on ElGamal encryption/decryption scheme. 

Furthermore, we integrated the two designs together and introduced a dual-circuit design that achieved a great improvement in area and power over a regular design that combines two IPs, one for ElGamal and another for CEG, without any resource sharing between them. Our architectures were implemented on a low-cost Xilinx Spartan-6 FPGA and area, delay, and power results were reported. 

%===============================================================================
\bibliographystyle{unsrt}
%\bibliographystyle{abbrv}
% argument is your BibTeX string definitions and bibliography database(s)
\bibliography{trojbib3}
%===============================================================================
% that's all folks
\end{document}